\documentclass[twocolumn,superscriptaddress,nofootinbib,preprintnumbers,showpacs,tightenlines,notitlepage]{revtex4}
\usepackage[latin1]{inputenc}
\usepackage{graphicx}
\usepackage{amssymb}
\usepackage{color}
\usepackage{float}
\usepackage{amsmath}
\usepackage{amsfonts}
\usepackage{dcolumn}
\usepackage{hyperref}
\usepackage{amsthm}
\usepackage{color}

\def\uudot{\dot{u}}
\def\3nab{\tilde{\nabla}}

\def\la {\langle}
\def\ra {\rangle}

\def\be {\begin{equation}}
\def\ee {\end{equation}}
\def\ba {\begin{eqnarray}}
\def\ea {\end{eqnarray}}

\newcommand{\bra}[1]{\left(#1\right)}
\newcommand{\bras}[1]{\left[#1\right]}
\newcommand{\brac}[1]{\left\{#1\right\}}

\newcommand{\sfr}[2]{{\textstyle\frac{#1}{#2}}}

\newcommand{\lc}{\varepsilon}
\newcommand{\E}{{\mathcal E}}

\newcommand{\barray}{\begin{array}}
\newcommand{\earray}{\end{array}}
\newcommand{\e}{e}
\newcommand{\N}{N}

\newcommand{\del}{\nabla}

 \newcommand{\nab}{\nabla}

\newcommand \ep {\epsilon}

\newcommand{\na}{\nabla}
\newcommand{\udot}{{\mathcal A}}
\begin{document}

\title{Astrophysical Black Hole horizons in a cosmological context: Nature and possible consequences on Hawking Radiation}
\author{George F R Ellis}
\email{George.Ellis@uct.ac.za}
\affiliation{Mathematics Department and ACGC, University of Cape Town, Rondebosch, 7701 South Africa}
 \author{Rituparno Goswami}
\email{Goswami@ukzn.ac.za}
\affiliation{Astrophysics and Cosmology Research Unit, School of Mathematics, Statistics and Computer Science, University of KwaZulu-Natal, Private Bag X54001, Durban 4000, South Africa.}
\author{Aymen I. M. Hamid}
\email{aymanimh@gmail.com\\ Permanent Address: Physics Department, University of Khartoum, Sudan.}
\affiliation{Astrophysics and Cosmology Research Unit, School of Mathematics, Statistics and Computer Science, University of KwaZulu-Natal, Private Bag X54001, Durban 4000, South Africa.}\author{Sunil D. Maharaj}
\email{Maharaj@ukzn.ac.za}
\affiliation{Astrophysics and Cosmology Research Unit, School of Mathematics, Statistics and Computer Science, University of KwaZulu-Natal, Private Bag X54001, Durban 4000, South Africa.}

\begin{abstract}
This paper considers the nature of apparent horizons for astrophysical black holes situated in a realistic cosmological context. Using semi-tetrad covariant methods we study the local evolution of the boundaries of the trapped region in the spacetime. For a collapsing massive star immersed in a cosmology with Cosmic Background Radiation (CBR), we show that the initial 2-dimensional marginally trapped surface bifurcates into inner and outer horizons. The inner horizon is timelike while the continuous CBR influx into the black hole makes the outer horizon spacelike. We discuss the possible consequences of these features for Hawking radiation in realistic astrophysical contexts.
\end{abstract}


\maketitle
 
\section{Introduction}
\label{sec:intro}

This paper is about the location of apparent horizons when astrophysical black holes form from the collapse of a single astrophysical object in a cosmological context. A current view with many proponents  is that when black hole formation takes place, a singularity forms and is hidden by an event horizon, but then the blackhole completely evaporates away within a finite time due to Hawking radiation \cite{Haw74,BirDav84,HawPen96}.The outcome  however has recently been challenged by various people  proposing that the black hole will not evaporate away \cite{Aharanov,Gidding,Strom,Ell13}, while others propose, using different arguments, that a global event horizon will never form \cite{Mer14,Bar14}. 

A key issue in this debate is the location of local apparent horizons, characterised as marginally outer trapped 3-surfaces (MOTS), which are arguably crucial to Hawking radiation emission. MOTS are important features of the geometry of black holes, inter alia leading to the prediction of the existence of singularities in the classical black hole case \cite{Pen65,HawEll73}. They are distinct from the global event horizon in realistic dynamical cases, when a pair of MOTS surfaces form (an inner one - the IMOTS and an outer - the OMOTS) as the radius of the collapsing object falls in past the critical radius $r = 2M$ (see Fig.1 and Fig. 2 and their descriptions).
They will be affected by infalling cosmic background radiation (CBR), as well as by backreaction from ingoing and outgoing Hawking radiation; and these effects must be taken  into account in realistic models of black hole radiation. 

In this paper we will show under what conditions each of these surfaces is timelike or spacelike, and where they are located. We find that the effect of the CBR on the nature of the outer MOTS surface may be a crucial feature in determining the context of Hawking radiation. Also, in realistic astrophysical contexts, the inner surface does not lie where Bardeen places it in his recent paper on black hole existence \cite{Bar14}; this may significantly affect the outcomes he claims.
     
A further paper will explore the implications for emission of Hawking radiation and black hole evaporation. This paper only considers spherically symmetric collapse; the situation will be much more complex in the case of rotating black holes, but the spherical case indicates the kind of outcomes we may expect in that case also. 

\subsection{The context} 
This section sets the general context within which the issue of the location of the event horizon arises. We are concerned with the case of astrophysical black holes situated in a realistic cosmological context. There are then three forks in the possibilities that will determine the outcome of Hawking radiation emission processes. The first two possibilities motivate the further studies in this paper.

\subsubsection{The cosmic context}
Considering black hole formation in a cosmological context, two features need to be taken into account, based on our present understanding of the universe.\\

\textbf{Cosmic Background Radiation}. Since the time of decoupling, the universe is everywhere pervaded by cosmic background radiation (CBR), initially at a temperature of 4000K but then cooling down to the present 2.73K, when the radiation is microwave background radiation  (CMB) \cite{PetUza09,EllMaaMac12}.  

Consequently any black hole formation will be subject to this infalling radiation \cite{Wieetal14}. This will affect the nature of the OMOTS surface, as discussed below.\\

\textbf{Cosmic constant}. The late universe is dominated by dark energy that is consistent with existence of a small positive cosmological constant. As the cosmological constant does not decay away, the late time evolution of the universe is de Sitter, and future infinity is spacelike \cite{Pen64,HawEll73}. 

This means that, unlike the asymptotically flat case, there is no single event horizon for any observer outside a black hole. Different cosmological event horizons (which are in the far future) occur for different observer motions that end up at different places $p_i$ on future infinity (see Figure 4 below), because the event horizon for an observer ending up at $p_i$ is, by definition, the past light cone of $p_i$ \cite{Pen64,HawEll73}. 

\subsubsection{Key question 1: Local of global?}

The first key question is whether the location of Hawking radiation emission is locally or globally determined. Is it just outside a globally determined event horizon, as argued strongly by Don Page (private communication), or just outside a local horizon (a MOTS surface) that is locally determined, as argued strongly by Visser \cite{Vis01} and supported by others \cite{Nie08,Cli08,ParWil00}. { In fact the source of the Hawking radiation is still a grey area and we need to perform very detailed and deep semiclassical investigations on the global aspects of physically realistic gravitational collapse to pinpoint such a source. The attempts so far, to introduce a timelike emitting surface 
outside the event horizon (for example \cite{His81, Suss, Hay06, Bar14}) are ad-hoc in the sense that the existence of such a surface lacks a proper geometrical interpretation. In the last two works mentioned above, this emitting surface has zero energy density but non-zero surface stress which is unphysical. It is hard to motivate why such a surface should exist in an absolutely regular and future asymptotically simple spacetime outside the black hole. On the other hand trapped regions (which have proper geometrical interpretation in terms of  null congruences) do play an important role in the ``particle pair-creation picture'' of Hawking radiation. Hence for a local analysis it is quite natural to assume that the boundary of the trapped region is the source of the radiation. In the case of an unperturbed Schwarzschild black hole, since this boundary exactly coincides with the future event horizon, the event horizon and it's vicinity can be considered as a global source. However the problem arises when this degeneracy is broken by infalling matter in the black hole.
}
The viewpoint underlying the basis of this paper is that the Hawking radiation emission must be locally, rather than globally, determined.  This is for two reasons. Firstly, this seems to be the only way the concept of a blackhole locally emitting radiation makes sense (it seems absurd that we have to wait until the end of the universe before we know what happens locally \cite{Vis01}). Secondly, because of the point just made: in a realistic context, future infinity outside the black hole is spacelike rather than null, so different observers who are initially near the black hole will experience different event horizons. Which one should we associate with the radiation? The prescription is not well defined.  

\subsubsection{Key question 2: Timelike or spacelike?}
Assuming that Hawking radiation emission is associated with a MOTS surface, the question then is, does it matter whether this surface is timelike, spacelike, or null? If the calculation is based on the idea of tunnelling (e.g. \cite{ParWil00}) then it will be applicable only if the MOTS surface is timelike or null (when ``inside'' and ``outside'' can be defined). If it is spacelike, then a tunnelling viewpoint is simply not applicable (spacetime regions will be ``before'' or ``after'' the surface, but not inside or outside it). The same applies to heuristic explanations based on one of a pair of virtual particles being trapped behind the horizon (they cannot be trapped behind a spacelike surface)  or using S-matrix ideas (scattering takes place off a timelike world tube, not a spacelike surface). These mechanisms can only work if the horizon is timelike.\\

Now the particle picture is considered suspect by many workers. As emphasized by Paul Davies and Malcolm Perry (private communication) what is needed to make this conclusive is to calculate the stress energy tensor associated with local horizons (see e.g. \cite{BirDav84,Cli08}),  and see if this confirms what is suggested by the particle picture or not. That will be the subject of a separate paper. 

For the present paper, the issue is that whether a MOTS surface is timelike or spacelike is not only geometrically important, because it determines the relation of the MOTS surface to global event horizons, but it may also play a key role in the Hawking radiation emission.   

\subsubsection{Key question 3: vacuum or fluid?}
The above two questions assume that the emission of radiation due to quantum processes in a collapsing black hole context is due to the properties of the vacuum domain near the event horizon or trapping surface. However there is another effect at work: the interior of the collapsing fluid is a time-varying gravitational field due to the collapse process, and this can potentially lead to particle creation in an evolving fluid filled spacetime, as noted long ago by Leonard Parker \cite{Par69}.  This is the mechanism of particle creation by a fluid collapsing as it forms a black hole that is discussed by Hawking (\cite{Haw73}: pp 207-208) and Birrell and Davies (\cite{BirDav84}: pp 250-262). In this case the role of the horizon is not creation of the  radiation, rather it is modulation of the propagation of the radiation that has already been emitted in the time-dependent gravitational field of the fluid; the result is that the outcome is independent of the details of the fluid collapse \cite{Haw73,BirDav84}. 

This is also the mechanism considered by Mersini-Houghton in her recent paper \cite{Mer14}, based on the Hartle-Hawking vacuum, and the assumption that the energy momentum tensor of the ingoing Hawking radiation is that of radiation in thermal equilibrium. However the Hartle-Hawking vacuum is based on the unphysical case of a white hole being in thermal equilibrium with the black hole; also the ingoing radiation may have a form different from thermal equilibrium.  Thus it is still an open question as to how significant this mechanism is compared to the horizon based mechanisms usually associated with Hawking radiation \cite{Haw73}. 

What we need to do to convincingly determine the outcome of this fluid based mechanism is to study the evolution of a collapsing body, and then ask what difference do quantum fields make to this dynamical situation. It will be significant to see how this relates to the nature and location of global and local horizons that are discussed in this paper. 
 
\subsection{The dynamical horizons}
There is a marginal outer trapped 3-surface (the OMOTS) that
lies outside the outgoing initial null surface
generated by the initial marginally trapped 2-surface (the
$S_{MOTS}$) that marks the onset of black hole dynamics. The OMOTS
surface is locally determined; it moves outwards with time
because of incoming Cosmic Microwave Background radiation, so that the
associated mass $m_{out}(u)$ increases with time, and the surface $r
= 2M_{out}(u)$ is spacelike. Because the OMOTS bounds the trapped domain in
spacetime (the outward directed null geodesics are converging inside this surface), it determines the outside edges of any future
singularity that may occur. Thus it non-locally determines the location of the
event horizon. OMOTS is a spacelike dynamical horizon whose properties
characterise the exterior mass
$m_{out}$ and angular momentum $J_{out}$ of the body. However its nature could possibly be changed to timelike at late times by the back reaction of Hawking radiation on its geometry. That is one of the important issue to be investigated.

We make the case below that there is an inner timelike marginally outer trapped 3-surface (the IMOTS),
which is a dynamical horizon \cite{AshKri02}. This surface lies inside both the OMOTS surface and the event horizon and it's
location is locally determined. It is timelike and also emanates from the 
$S_{MOTS}$ 2-surface, which is therefore a bifurcation surface originating both MOTS surfaces. It is potentially possible that backreaction from Hawking radiation causes them to merge again in the future at a final $F_{MOTS}$ 2-surface, if the OMOTS surface eventually becomes timelike. Whether that happens or not depends on detailed balance between local dynamics of the fluid and the incoming Hawking radiation in those cases where the OMOTS is timelike at late times. 

\subsection{This paper}
The paper considers these issues in the case of a single black hole with
spherical symmetry, where the relevant exterior solution is the exterior
Schwarzschild solution surrounding the single collapsing mass. The result is
probably stable in the case of more general geometries such as rotating
black holes, perturbed black holes, and if there are later infalling shells
of matter. The paper does not consider multiple black holes or black hole
collisions. Since we consider only astrophysically relevant situations,
it also does not consider charged black holes either.

As we confine our attention to spherically symmetric black holes produced by the gravitational collapse of a 
massive star, the Schwarzschild solution is the basic relevant exterior metric but the interior will be modelled by spherical fluid body: it might be a Friedmann-Lema\^{i}tre-Robertson-Walker (FLRW) metric, a Lema\^{i}tre-Tolman-Bondi (LTB) metric, or a spherical metric with pressure \cite{EllMaaMac12}. However in addition there may be incoming or outgoing radiation in both the exterior and interior regions; we therefore need to consider generic spherically symmetric metrics. For technical reasons it is convenient to consider a class of spacetimes which are a generalisation of spherically symmetric metrics: namely {\it Locally Rotationally Symmetric} (LRS) class II spacetimes \cite{EllisLRS,Stewart-Ellis,vanEll96}.  These are evolving and vorticity free spacetimes with a 1-dimensional isotropy group of spatial rotations at every point. Except for  few higher symmetry cases, these spacetimes have locally (at each point) a unique preferred spatial direction that is covariantly defined.

To describe this class of spacetimes in terms of metric components, we use the most general line element for LRS-II can be written as \cite{Stewart-Ellis}
\ba
   ds^2 &=& -A^{2}(t,\chi)\,dt^2+B^2(t,\chi)\,d\chi^2 \nonumber\\
   &&+C^2(t,\chi)\,[\,dy^2+D^2(y,k)\,dz^2\,] \;,\label{LRSds}
\ea
where $t$ and $\chi$ are parameters along the integral curves of the timelike vector field  $u^a = A^{-1}\delta^a_0$ and the preferred spacelike vector field $e^a = B^{-1}\delta^a_\nu$. The function $D(y,k)= \sin y,y, \sinh y$ for $k=(1,0,-1)$ respectively. The 2-metric $\,dy^2+D^2(y,k)\,dz^2$ describes spherical, flat, or open homogeneous and isotropic 2-surfaces for $k=(1,0,-1)$. Spherically symmetric spacetimes are the $k=1$ subclass of LRS-II spacetimes.  

We can easily see that the physically interesting spherically symmetric spacetimes (for example the Schwarzschild, FLRW, LTB, and Vaidya  solutions) fall in the class LRS-II. Hence both the interior and the exterior of a collapsing star can be described by this class, which can be characterised covariantly through the properties of the vector fields $u^a $ and $e^a$ \cite{vanEll96}. A parallel paper is being developed by Hellaby to study the same problem using a metric based formalism. The basic results presented in this paper are supported by studies using that different formalism.

Unless otherwise specified, we use natural units ($c=8\pi G=1$) throughout this paper, Latin indices run from 0 to 3. 
The symbol $\nabla$ represents the usual covariant derivative and $\partial$ corresponds to partial differentiation. 
We use the $(-,+,+,+)$ signature.  

\section{Covariant description of LRS-II spacetimes}

In this section we give a brief summary of the semi-tetrad 1+1+2 covariant description of LRS-II spacetimes. For more details we refer to \cite{EllisCovariant,vanEll96,Clarkson:2002jz,Betschart:2004uu,Clarkson:2007yp}. 

As the first step of the covariant description \cite{Ell71}, we define a time-like congruence with a unit tangent vector $u^a$. One natural and obvious 
choice of this vector may be the tangent to the matter flow lines. Any
4-vector $X^a$ in the manifold can then be projected onto the 3-space by the projection
tensor $h^a{}_b=g^a{}_b+u^au_b$ as $X^a=Xu^a+X^{\la a\ra}$, where $X$ is the scalar along $u^a$ and $ X^{\la a\ra}$ is the projected 3-vector.
A natural definition for two kinds of derivatives are then: 
\begin{itemize}
\item The \textit{covariant time derivative} along the observers' worldlines (denoted by a dot). Hence for any tensor  $ S^{a..b}{}_{c..d}$ we have 
$\dot{S}^{a..b}{}_{c..d}{} \equiv u^{e} \nab_{e} {S}^{a..b}{}_{c..d} $.
\item Fully orthogonally \textit{projected covariant derivative} $D$ with the tensor $ h_{ab} $,  with total projection on all the free indices. Hence we have 
$ D_{e}S^{a..b}{}_{c..d}{} \equiv h^a{}_f
h^p{}_c...h^b{}_g h^q{}_d h^r{}_e \nab_{r} {S}^{f..g}{}_{p..q}$.
\end{itemize}
This 1+3 splitting defines the 3-volume element naturally as $\ep_{a b c}=-\sqrt{|g|}\delta^0_{\left[ a\right. }\delta^1_b\delta^2_c\delta^3_{\left. d \right] }u^d$. 
For the LRS-II class of spacetimes, the covariant derivative of the timelike vector $u^a$ can be irreducibly split in the following way
\be
 \nabla_au_b=-A_au_b+\frac13h_{ab}\Theta+\sigma_{ab},
 \ee 
where $A_a=\dot{u_a}$ is the acceleration,
$\Theta=D_au^a$ is the expansion, $\sigma_{ab}=D_{\la a}u_{b \ra}$ (where the angle brackets denote the projected symmetric trace-free part)
is the shear tensor. Also for these spacetimes all the independent components of the Weyl tensor can be expressed in terms of the 
projected symmetric trace free gravito-electric tensor 
defined as $E_{ab}=C_{acbd}u^cu^d=E_{\la ab\ra}$. The gravito-magnetic part of Weyl tensor \cite{Maartens:1997fg} is identically zero.
Using this timelike vector $u^a$, the energy momentum tensor for a general matter field is decomposed as follows
\be
T_{ab}=\rho u_au_b+q_au_b+q_bu_a+ph_{ab}+\pi_{ab}\;,
\ee
where $\rho=T_{ab}u^au^b$ is the energy density, $p=(1/3 )h^{ab}T_{ab}$ is the isotropic pressure, $q_a=q_{\la a\ra}=-h^{c}{}_aT_{cd}u^d$ is the 3-vector defining 
the heat flux and $\pi_{ab}=\pi_{\la ab\ra}$ is the anisotropic stress. 

The symmetry of the LRS-II spacetimes, having a preferred spatial direction, points towards an extension of the 1+3 splitting mentioned above, by further splitting the 3-space 
by the  preferred spatial vector $e^a$. This is commonly known as 1+1+2 splitting. This allows us to derive a set of covariant scalar variables which are more advantageous to treat these systems. This spatial vector is  orthogonal to $u^a$ such that it satisfies
$\e^a\e_a=1,~u^a\e_a=0$. The 1+3 projection tensor
$h_{a}^{~b}\equiv g_{a}^{~b}+u_au^b$ combined with $e^a$ defines a new projection tensor defined as $ \N_a^{~b}\equiv g_{a}^{~b}+u_au^b-\e_a\e^b$ which projects
vectors orthogonal to $\e^a$ and $u^a$
($\e^a\N_{ab}=0=u^a\N_{ab}$) onto a 2-surface which is defined as the sheet ($\N_a^{~a}=2$). The volume element of this 
2-surface is the Levi-Civita 2-tensor $\lc_{ab}\equiv\ep_{abc}\e^c = u^d\eta_{dabc}e^c$. The preferred spatial vector $e^a$ introduces two new derivatives as a 
natural result of the new splitting of the 3-space, for any 3-tensor $ \psi_{a...b}{}^{c...d} $ :
\begin{itemize}
\item The {\it hat derivative} is the spatial derivative along the vector $e^a$. Thus we have $\hat{\psi}_{a..b}{}^{c..d} \equiv e^{f}D_{f}\psi_{a..b}{}^{c..d}$.
\item The {\it delta derivative} is the projected spatial derivative on the 2-sheet by the projection tensor $N_a^{~b}$ and projected on all the free indices. Hence 
$\delta_f\psi_{a..b}{}^{c..d} \equiv N_{a}{}^{f}...N_{b}{}^gN_{h}{}^{c}..
N_{i}{}^{d}N_f{}^jD_j\psi_{f..g}{}^{i..j}$.
\end{itemize}

The set of 1+1+2 covariant scalars that fully describe LRS-II spacetimes are 
$\brac{\udot, \Theta,\phi, \Sigma,\E, \rho, p, \Pi, Q }$, which are defined as follows
\ba
\uudot^a=\udot \e^a\;,&&\phi=\delta_ae^a\;,\\
 \sigma_{ab}=\Sigma\bra{\e_a\e_b-\sfr{1}{2}\N_{ab}} \;,&&E_{ab}={\cal E}\bra{\e_a\e_b-\sfr{1}{2}\N_{ab}}\;,\\
 \pi_{ab}=\Pi\bra{\e_a\e_b-\sfr{1}{2}\N_{ab}}\;,&&q^a=Q \e^a\;.
 \ea
 And the full covariant derivative of $\e_a$ and $u_a$ is now written as
\ba\label{del_e}
\,\,\, \del_a\e_b&=&-\udot u_au_b +\bra{\Sigma+\sfr13\Theta}\e_a u_b+\sfr{1}{2}\phi\N_{ab},
\ea
\ba\label{del_u}
\del_au_b&=&
-\udot u_a\e_b+\e_a\e_b\bra{\sfr13\Theta+\Sigma}+\N_{ab}\bra{\sfr13\Theta-\sfr12\Sigma}\;.
\ea
We also write the useful relation
\be
\hat{u}_a =
\bra{\sfr13\Theta+\Sigma}\e_a \;.
\label{uhat}
\ee

 We now derive the equations governing the evolution and propagation of the covariant scalars that fully describes the LRS-II spacetimes. Using the Ricci identities for the vectors $u^a$ and $e^a$ and the doubly contracted Bianchi identities, we obtain the following:
 \begin{widetext}
 
\begin{itemize}
\item\textit{Propagation}:
\ba
\hat\phi  &=&-\sfr12\phi^2+\bra{\sfr13\Theta+\Sigma}\bra{\sfr23\Theta-\Sigma}-\sfr23\bra{\rho+\Lambda}
    -\E -\sfr12\Pi,\,\label{hatphinl}
\\  
\hat\Sigma-\sfr23\hat\Theta&=&-\sfr32\phi\Sigma-Q\
,\label{Sigthetahat}
 \\  
\hat\E-\sfr13\hat\rho+\sfr12\hat\Pi&=&
    -\sfr32\phi\bra{\E+\sfr12\Pi}
    +\bra{\sfr12\Sigma-\sfr13\Theta}Q.\;\label{hateps}
\ea

\item\textit{Evolution}:
\ba
   \dot\phi &=& -\bra{\Sigma-\sfr23\Theta}\bra{\udot-\sfr12\phi}
+Q\ , \label{phidot}
\\   
\dot\Sigma-\sfr23\dot\Theta  &=&
-\udot\phi+2\bra{\sfr13\Theta-\sfr12\Sigma}^2
        +\sfr13\bra{\rho+3p-2\Lambda}-\E+\sfr12\Pi\, ,\label{Sigthetadot}
\\  
\dot\E -\sfr13\dot \rho+\sfr12\dot\Pi &=&
    +\bra{\sfr32\Sigma-\Theta}\E
    +\sfr14\bra{\Sigma-\sfr23\Theta}\Pi
   +\sfr12\phi Q-\sfr12\bra{\rho+p}\bra{\Sigma-\sfr23\Theta}\ . \label{edot}
\ea

\item\textit{Propagation/evolution}:
\ba
   \hat\udot-\dot\Theta&=&-\bra{\udot+\phi}\udot+\sfr13\Theta^2
    +\sfr32\Sigma^2 
    +\sfr12\bra{\rho+3p-2\Lambda}\ ,\label{Raychaudhuri}
\\
    \dot\rho+\hat Q&=&-\Theta\bra{\rho+p}-\bra{\phi+2\udot}Q -
    \sfr32\Sigma\Pi,\,
\\    \label{Qhat}
\dot Q+\hat
p+\hat\Pi&=&-\bra{\sfr32\phi+\udot}\Pi-\bra{\sfr43\Theta+\Sigma} Q
    -\bra{\rho+p}\udot\ .
\ea
\end{itemize}
\end{widetext}
The 3-Ricci scalar of the spacelike 3-space orthogonal to $u^a$ can be expressed as
\be
^3R =-2\bras{\hat \phi +\sfr34\phi^2 -K}\;,\label{3sca}
\ee
where $K$ is the Gaussian curvature of the 2-sheet defined by
$^2R_{ab}=KN_{ab}$. In terms of the covariant scalars we can write the Gaussian curvature $K$ as
\be
K = \sfr13\bra{\rho+\Lambda}-\E-\sfr12\Pi +\sfr14\phi^2
-\bra{\sfr13\Theta-\sfr12\Sigma}^2\ . \label{gauss}
\ee
Finally the evolution and propagation equations for the Gaussian curvature $K$ are
\ba
\dot K = -\bra{\sfr23\Theta-\Sigma}K\ , &&\hat K = -\phi K\ . \label{Kdothat}
\ea

\section{Null geodesics in LRS-II spacetimes}

On any spacetime $(\mathcal{M},g)$, null geodesics (light rays) are characterised by the curves $x^{a}(\nu)$, where $\nu$ is an affine parameter along the geodesics. 
The tangent to these curves is defined by $k^{a} = \sfr{dx^{a}(\nu)}{d\nu}$, where $k^a$ is a null vector obeying $k^{a}k_{a} = 0$. 
Also, since the tangent vector to the geodesic is parallely propagated to itself, we can write
\be\label{geo-k}
k^{b}\na_{b}k^{a} =\frac{\delta k^{a}}{\delta \nu} = 0~, 
\ee
where $\sfr{\delta }{\delta \nu} = k^{b}\na_{b}$ as the derivative along the ray with respect to the affine parameter.

On LRS-II spacetimes if light rays moves along the preferred spatial direction (for the special case of spherical symmetry this is equivalent to radial null rays), then the sheet components of these null curves are zero. Also at this point we may define the notion of locally {\it outgoing} and {\it incoming} null geodesics with respect to the preferred spatial direction. Consider any open subset ${\mathcal S}$ of $(\mathcal{M},g)$ and let $x^{a}(\nu)$ be a null geodesic in ${\mathcal S}$. Let $k^a$ be the tangent to this geodesic. 
If $e^ak_a>0$ in ${\mathcal S}$ then the geodesic is considered to be outgoing with respect to the preferred direction in ${\mathcal S}$. Similarly 
$e^ak_a<0$ denotes an incoming geodesic. Therefore for LRS-II spacetimes, the equation of the tangent to the outgoing null geodesics along the preferred spatial direction 
can be written as 
\be
k^a=\frac{E}{\sqrt{2}}\bra{u^a+e^a}\;,
\ee
where $E:[u,e]\rightarrow\mathbb{R}$ denotes the energy of the light ray. The propagation of $E$ along the null geodesic is given by the geodesic equation (\ref{geo-k}) as
(using (\ref{del_e}) and (\ref{del_u}))
\be
E'\equiv\frac{\delta E}{\delta \nu} = \mp E^{2}\udot - E^{2} \bra{\Sigma+\sfr13\Theta}~.\label{Eprime2}
\ee
We can easily see that the hypersurface orthogonal to null vector $k^a$, contains $k^a$. Therefore we need a different construction to define a locally orthogonal space with respect to the outgoing null geodesics.
Let us now define the projection tensor $\tilde{h}_{ab}$, which projects tensors and vectors into the 2-D 
screen space orthogonal to $k^a$, as \cite{deSwardt:2010nf}
\be
\tilde{h}_{ab}\equiv g_{ab}+2k_{(a}l_{b)}, \,\,\tilde{h}^a_a=2,\,\,\tilde{h}_{ac}\tilde{h}^c_b=\tilde{h}_{ab},\,\,\tilde{h}_{ab}k^b=0,\label{ht}
\ee
where $l_a$ is the null ingoing geodesic that obeys
\be
l^al_a=0,\,\,k^al_a=-1\,\, \text{and} \,\,\frac{\delta l^a}{\delta \nu}=k^b\nabla_bl^a=0.
\ee
Using these definitions, the general form of $l^a$ can be written as
\be
l^a=\frac{1}{\sqrt{2}E}\bra{u^a-e^a}\label{l},
\ee
and substituting (\ref{l}) into (\ref{ht}) the screen-space projection tensor is obtained as
\be
\tilde{h}_{ab}=g_{ab}+u_au_b-e_ae_b=N_{ab}
\ee
Hence we note that the 2-D screen projection tensor is absolutely same as the 2-sheet projection tensor $N_{ab}$.
For the LRS-II spacetimes the 1+3 decomposition of the covariant derivative of the null vector $k^a$ 
is given as \cite{deSwardt:2010nf}
\be
\nabla_bk_a=\sfr12 \tilde{h}_{ab}\tilde{\Theta}_{out}+\tilde{\sigma}_{ab}+\tilde{X}_ak_b+\tilde{Y}_bk_a
+\lambda k_ak_b,
\ee
where 
\be
\tilde{X}_a=\frac1Ee^d\nabla_dk_a,\tilde{Y}_a=\frac1Ee^d\nabla_ak_d,\lambda=-\frac{1}{E^2}e^ce^d\nabla_dk_c,
\ee
and  $\tilde{\Theta}_{out}$ and $\tilde{\sigma}_{ab}$ represent the expansion and shear of 
of the outgoing null congruence respectively. From the above definitions and the geometry of LRS-II spacetimes we can easily calculate that 
\ba
\tilde{\Theta}_{out}\equiv\tilde{h}^{ab}\nabla_bk_a
&=&\frac{E}{\sqrt{2}}N^{ab}\nabla_a\bra{u_b+e_b}
 \label{tth}.
\ea
Now using (\ref{del_e}) and (\ref{del_u}) in (\ref{tth}) we obtain,
\be\label{thetaout}
\tilde{\Theta}_{out}=\frac{E}{\sqrt{2}}\bra{\sfr23\Theta-\Sigma+\phi}\;.
\ee
A similar 1+3 decomposition of the covariant derivative of the ingoing null geodesic $l^a$ can be performed, giving 
\be\label{thetain}
\tilde{\Theta}_{in}=\frac{1}{\sqrt{2}E}\bra{\sfr23\Theta-\Sigma-\phi}\;.
\ee

\section {Marginally outer trapped surfaces}

As described in \cite{HawEll73}, let us consider a spherical emitter of light, surrounding a massive body, emitting a flash of light. In a normal situation, (like Minkowski spacetime or Schwarzschild spacetime outside the horizon) the volume expansion of the outgoing null congruence orthogonal to the sphere is always positive ($\tilde{\Theta}_{out}>0$) 
while that of the incoming congruence is always negative ($\tilde{\Theta}_{in}<0$). However, if a sufficiently large amount of matter is present within the emitting sphere, the volume expansion of the outgoing null congruence orthogonal to the sphere becomes negative. This is the case where both outgoing and ingoing wavefronts collapse towards the centre of   the emitting sphere. In this case, the emitting sphere is then called a {\em Closed trapped 2-surface}.  The collection of all closed trapped 2-surfaces in a four dimensional manifold constitutes a four dimensional trapped region. {\it Marginally outer trapped surfaces} (MOTS) are the 3 dimensional boundary of the trapped region. We will now derive the equations for a MOTS surface and it's evolution in the $[e,u]$ plane. Similar derivations were done in \cite{Booetal05} using a different formalism.

\subsection{Equation for MOTS is LRS-II spacetimes}

To define MOTS we generalise the definition of dynamical horizons by Ashtekar and
Krishnan \cite{AshKri02} by removing the restriction that it is a
spacelike surface. Thus,
\begin{quote}
\textbf{Definition: Marginally Outer Trapped Surface (MOTS):} A  smooth, three-dimensional
sub-manifold $H$
in a spacetime $(\mathcal{M},g)$ is said to be a \textit{Marginally outer trapped surface} if it is foliated by a preferred family of
2-spheres such that, on each leaf $S$, the expansion $\tilde{\Theta}_{out}$ of the outgoing null normal $k^a$ vanishes and the expansion
$\tilde{\Theta}_{in}$ of the other ingoing null normal $l^a$ is strictly
negative.
\end{quote}
 Consequently a MOTS is a 3-manifold which is foliated by marginally trapped 2-spheres. On this definition, such 3-surfaces can be timelike, spacelike, or null, with rather different properties. The essential point is that they are locally defined, and therefore are able respond to local
dynamic change. We don't need to know what is happening at infinity
in order to determine a local physical effect.

At this point let us consider spacetimes that have a positive sheet expansion, $\phi\equiv\delta_a\e^a>0$. In the geometrical context this implies that at any given epoch, the local Gaussian curvature of the two sheets are monotonically decreasing functions of the affine parameter along the preferred spatial direction. Schwarzschild, FLRW, and LTB spacetimes have this property. Furthermore, by the symmetries of LRS-II spacetimes, we can easily see that it suffices to study the one dimensional MOTS curve in the local $[u,e]$ plane to determine it's local properties. Now using equations (\ref{thetaout},\ref{thetain}) it is evident that the equation of the MOTS curve in the  local $[u,e]$ plane is given by
\be
\Psi\equiv \bra{\sfr23\Theta-\Sigma+\phi}=0\;.
\ee
Thus, if $\phi>0$, on the MOTS curve we have  $\tilde{\Theta}_{out}=0$ and $\tilde{\Theta}_{in}<0$, as it should be by the above definition. Here we would like to show another important geometrical property of MOTS. Let us calculate the quantity $\nabla_a K\nabla^aK$ for a spherically symmetric spacetime (where $K\ne 0$):
\be
\nabla_a K\nabla^aK=-\dot{K}^2+\hat{K}^2\label{Geq}.
\ee
Now using (\ref{Kdothat}) in (\ref{Geq}) we get
\be
\nabla_a K\nabla^aK=\bra{\sfr23\Theta-\Sigma+\phi}\bra{\sfr23\Theta-\Sigma-\phi}K^2\;.
\ee
Hence for any 2-sheet on the MOTS, the gradient of it's local Gaussian curvature is null. This property allows us to easily locate the MOTS in a given spacetime.

As an example let us consider the spherically symmetric vacuum spacetime. Then by Birkhoff's theorem \cite{HawEll73} we know that the spacetime has an extra symmetry, it is either static or spatially homogeneous. Let us consider the static exterior part. Existence of a timelike Killing vector implies that $\Theta=\Sigma=0$ \cite{GosEllis}. 
Thus the horizon is described by the curve $\phi=0$. This is the {\it event horizon} of the Schwarzschild spacetime. Indeed if we calculate $\phi$ from the LRS-II field equations in Schwarzschild coordinates, we get 
\be
\phi=\frac2r\sqrt{1-\frac{2m}{r}},
\ee
and $\phi=0$ corresponds to the event horizon at $r=2m$.

\subsection{Evolution of MOTS }

Let us now describe how the MOTS evolve geometrically. What are the local conditions in terms of geometry or the matter energy momentum tensor that determine whether MOTS will be locally timelike, spacelike or null? As we stated earlier, due to the symmetries of LRS-II spacetimes, it suffices  to study the behaviour of the MOTS curve in the local $[u,e]$ plane.
Let the vector $\Psi^a=\alpha u^a+\beta e^a$ be the tangent to the curve $\Psi=0$ in the local $[u,e]$ plane. 
Then we must have $\Psi^a \nabla_a\Psi=0$. Since we know that $\nabla_a\Psi=-\dot{\Psi}u_a+\hat{\Psi}e_a$, we can immediately see 
that the slope of the tangent to the MOTS on the local $[u,e]$ plane is given by $\sfr{\alpha}{\beta}=-\sfr{\hat{\Psi}}{\dot{\Psi}}$.
Now using this decomposition with the field equations (\ref{hatphinl}) to (\ref{Qhat}), we obtain 
\ba\label{normal}
\nabla_a\Psi&=&\bra{\sfr13(\rho+\Lambda)+(p-\Lambda)-{\cal E}+\sfr12 \Pi-Q}u_a\nonumber\\
&&+\bra{-\sfr23(\rho+\Lambda)-\sfr12\Pi-{\cal E}+Q}e_a,
\ea
and hence
\be\label{alph}
\frac{\alpha}{\beta}=\frac{\sfr23(\rho+\Lambda)+\sfr12\Pi+{\cal E}-Q}{-\sfr13(\rho+\Lambda)-(p-\Lambda)+{\cal E}-\sfr12 \Pi+Q}\;.
\ee
We now define the notion of {\it Future Outgoing (Ingoing)} in the following way. If $\sfr{\alpha}{\beta}>0$ then the MOTS is said to be future outgoing and if $\sfr{\alpha}{\beta}<0$ then the MOTS is said to be future ingoing. The nature of the MOTS in terms of it being timelike, spacelike or null can also be determined by the ratio $\sfr{\alpha}{\beta}$. We can easily see that \begin{equation}
\Psi^a\Psi_a=\beta^2(1-\sfr{\alpha^2}{\beta^2}).
\end{equation}
Therefore  $\frac{\alpha^2}{\beta^2}> 1 (< 1)$ denotes the MOTS
to be locally timelike  (spacelike). If $\sfr{\alpha^2}{\beta^2}=1$ (as in the case of vacuum spacetime,  Schwarzschild or Schwarzschild-de Sitter, where all the thermodynamical terms vanish), the MOTS are null. 

It is interesting to see how the local  matter thermodynamical quantities, along with the Weyl curvature entirely determine the nature of the dynamical horizon. In special cases, 
where the matter is a perfect fluid ($Q=\Pi=0$), then the condition for the MOTS being timelike is given by (assuming $\Lambda=0$)
\be 
\bra{\sfr13\rho-p+2{\cal E}}\bra{\rho+p}<0\;.
\ee
Thus if the weak energy condition is satisfied so that  $\rho+p>0$, then $\rho>(3p-6{\cal E})$ would ensure a timelike MOTS.

\section{Misner-Sharp mass for LRS-II spacetimes}

In this section, we derive the Misner-Sharp \cite{Misner:1964je} mass equation for LRS-II spacetimes in terms of the 1+1+2 kinematical
quantities. We know, for the metric (\ref{LRSds}), we can define the mass function as \cite{Stephani:2003ika}:
 \be
{\cal M}(r,t)=\frac{C}{2}\bra{k-\nab_a C\nab^a C}\;,
\ee
where $C$ represents the physical radius and $k=+1,0,-1$ represents closed, flat or open 3-space geometry. Here we are concentrating on spherically symmetric spacetimes and hence we will only consider the case $k=1$. Then we can write $C=\frac{1}{\sqrt{K}}$, where $K$ is the Gaussian curvature of the spherical 2-sheets. Then we obtain the expression of the mass as
\be
{\cal M}=\frac{1}{2\sqrt{K}}\bra{1-\frac{1}{4K^3}\nab_a K\nab^a K}\;.
\ee
Geometrically, the above expression gives the amount of mass enclosed within the spherical shell at a given value of affine parameter of the integral curves of $e^a$ at a given 
instant of time.  Using the 1+1+2 decomposition of the covariant derivative for LRS-II together with (\ref{gauss}, \ref{Kdothat}), the Misner-Sharp
mass takes the form
\be
{\cal M}=\frac{1}{2K^{3/2}}\bra{\frac{1}{3}(\rho+\Lambda)-\E-\frac{1}{2}\Pi}.
\label{mass}
\ee
We can easily see from this equation, even in the case of vacuum spacetimes the mass does not vanish due to the electric part
of Weyl Curvature $\cal E$, which plays the role of the mass source. 

As an example lets us consider the Schwarzschild static spacetime again. Since it is a vacuum spacetime the thermodynamical quantities vanish and in Schwarzschild coordinates we have 
\be
{\cal E}=-\frac{2m}{r^3}\;.
\ee
In this case we see that the Misner-Sharp mass is the same as the Schwarzschild mass. 
As discussed in the previous section, on the MOTS $\nab_a K\nab^a K=0$  \cite{Hamid:2014kza}. Therefore the mass enclosed within the MOTS at a given instant of time (which can be regarded as the local mass of the black hole) is given by
\be
{\cal M}_{BH}=\frac{1}{2\sqrt{K_{BH}}}\;,
\ee
 where $K_{BH}$ is the Gaussian curvature of the black hole surface at any given instant of time.

\section{Oppenheimer-Snyder-Datt Collapse}
This is the simplest model of black hole formation from the dynamical collapse of a massive star \cite{osd,datt}. The interior of the spherically symmetric star is considered to be dustlike and is described by a flat FLRW metric. The exterior of the star is spherically symmetric vacuum and hence, by Birkhoff's theorem,  is a static Schwarzschild spacetime. These two spacetimes are matched using Israel-Darmois matching conditions (the first and second fundamental forms) at the boundary of the collapsing star, which is described by a comoving radius $r_B$ in the interior spacetime. When the star collapses to the singularity at the centre the spacetime becomes Schwarzschild. Though this model is a simplified toy model, it is widely accepted that any realistic massive star would collapse to a black hole in a similar fashion \cite{Pen65}.

\begin{figure}[htb!!!!]
\includegraphics[width=3.5in]{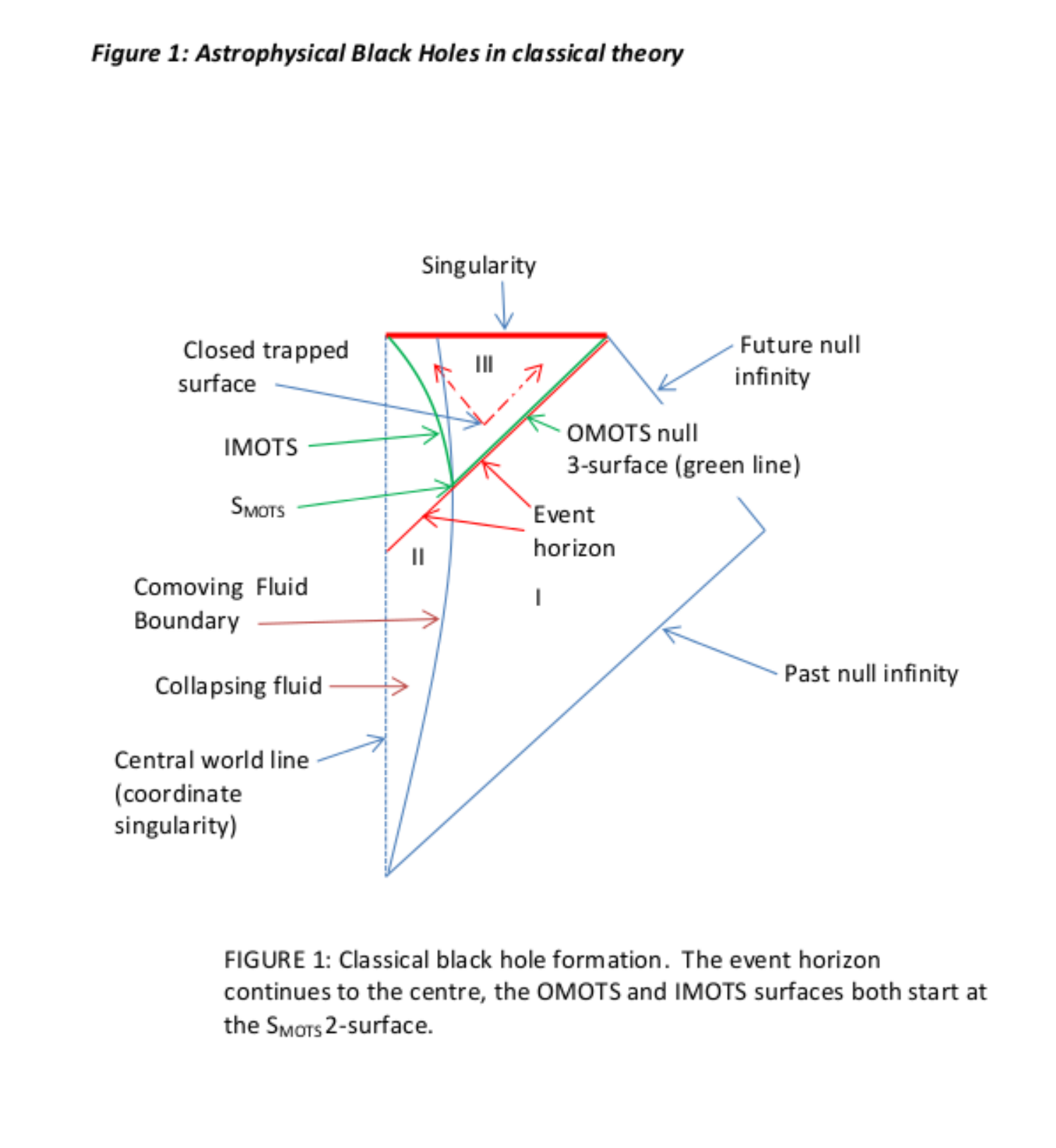}
\caption{The spacetime Diagram showing the formation of Event Horizon and Apparent Horizon (MOTS) for a spherically symmetric dust collapse.}
\label{Fig1}
\end{figure}
 From the Einstein's field equations one can explicitly derive the dynamics of the trapped region in the spacetime, which are as follows (see Figure 1) (see \cite{Joshibook1,Joshibook2,JoshiJVN} for discussions on the global evolutions of trapped regions):
\begin{itemize}
\item In the interior spacetime the boundary shell of the star gets trapped first. This is the instant of time when the radius of the star becomes equal to the Schwarzschild radius. The interior of the star is causally cut off from the exterior spacetime from this instant. As described in the previous section, at this instant the gradient of the Gaussian curvature of the 
boundary shell ($\nabla_aK_B$ ) becomes null. Thus a 2 dimensional initial MOTS is formed which we will denote as the $S_{MOTS}$. This initial $S_{MOTS}$ is described by a point in the $[u,e]$ plane, which is a bifurcation point. From this point two separate branches of 3-MOTS develop. The one which is in the exterior spacetime is called the {\it Outer} MOTS or OMOTS while the one which is in the interior of the star is called the {\it Inner} MOTS or IMOTS.
\end{itemize}
 To understand geometrically why the two MOTS surfaces exist, consider the domains where  $\tilde{\Theta}_{out}$ is positive and negative (Figure 2). The boundary between them is where $\tilde{\Theta}_{out}=0$. It is essentially a geometrical necessity that one cannot have a single starting trapped 2-sphere $S_{MOTS}$ with only one MOTS surface emanating from it. A single surface where $\tilde{\Theta}_{out} =0$ can't bound a domain where $\tilde{\Theta}_{out}<0$: there have to be two surfaces to make geometrical sense.
 
To understand physically why $\tilde{\Theta}_{out}=0$ on two surfaces, consider a sphere of radius $r$ round the centre  of the collapsing fluid containing a mass of fluid $m(r)$.  Close enough to the centre of the collapsing object, for small $r$, there is insufficient mass inside to cause refocusing: $m(r) < r/2$. That is why there is no refocussing close enough to the centre ( $\tilde{\Theta}_{out}>0$). The possibility of local trapping 2-spheres existing arises when $r$ is large enough that there is sufficient mass inside a sphere to cause refocusing: $m(r) > r/2$. The inner horizon IMOTS surface where $\tilde{\Theta}_{out}=0$  thus exists at any time $t$ at which there exists a first radius which includes sufficient matter to cause refocussing at that time (its like the refocussing of the past light cone in a FLRW model that occurs at $z=1.25$ in the Einstein de-Sitter case). The outer OMOTS horizon exists because outside the star there is vacuum, and the mass is then constant as the radius increases. So going to a sufficient radius  one will eventually  again find $r >2m$, and there is no refocussing outside the bounding surface where this first occurs. 
\begin{itemize}
\item Since the exterior spacetime is vacuum, all the matter thermodynamic terms are identically zero. Hence we have $\sfr{\alpha}{\beta}=1$. Consequently, the OMOTS is future outgoing and null. This exactly coincides with the event horizon of Schwarzschild spacetime.
\item The interior spacetime has flat FLRW geometry with dust like matter. Hence apart from the energy density $\rho$ all the other thermodynamic terms vanish and also the Weyl scalar is zero. Hence we get $\sfr{\alpha}{\beta}=-2$. Thus the IMOTS is a future ingoing timelike 3-surface. This surface reaches the centre of the collapsing star at the singularity.
\end{itemize}
The trapping region III and its boundaries separating it from (non-trapped) regions I and II are show in Figure 2. It is clear from this characterisation of the domains where $\theta_+$ is greater and less that zero that the MOTS surfaces have to be created as a pair at the $S_{MOTS}$ 2-sphere.

\begin{figure}[htb!!!!]
\includegraphics[width=3.5in]{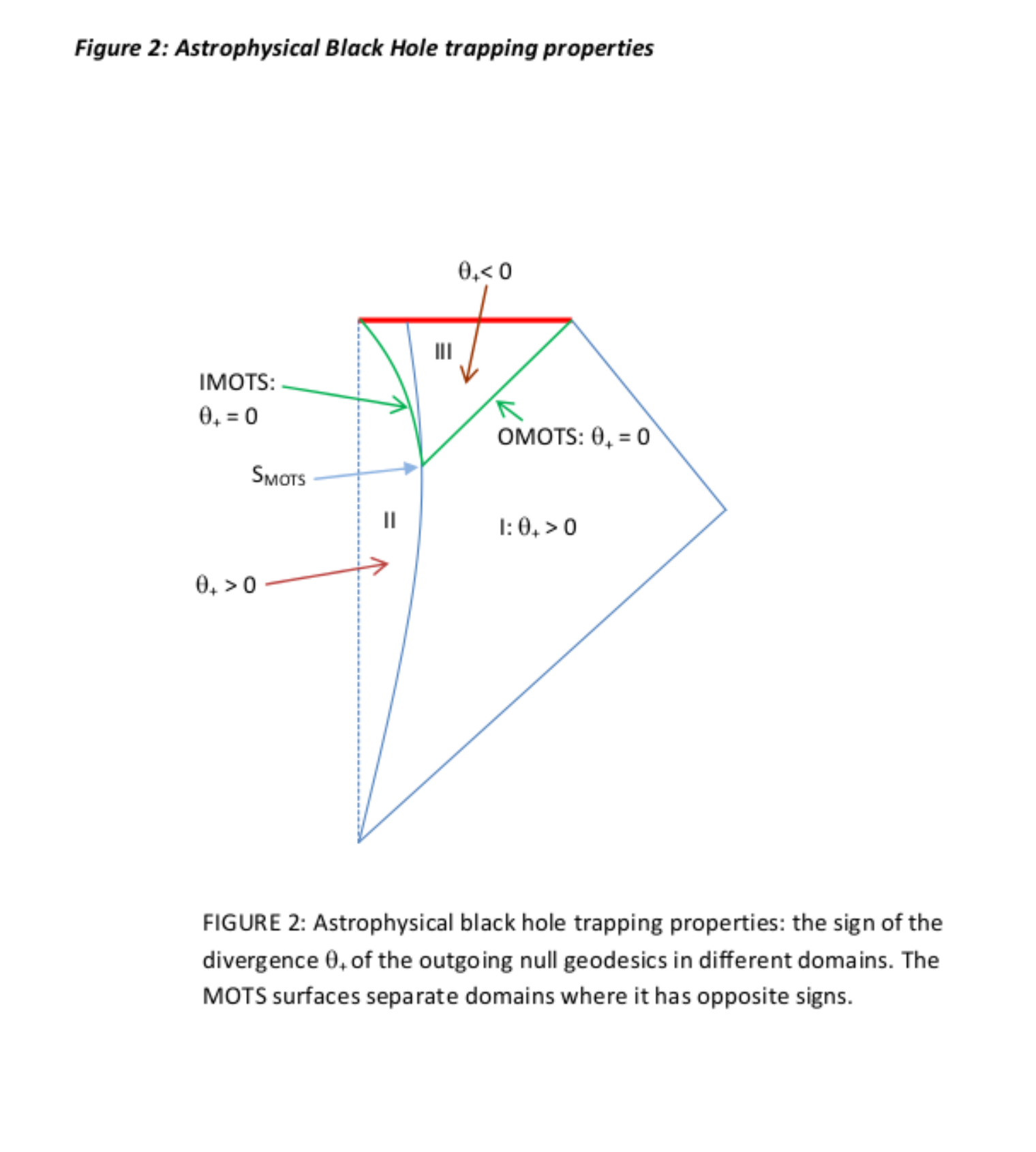}
\label{Fig2}
\end{figure}

A Ricci tensor singularity will occur inside the fluid as the fluid
collapses and $K_B(\tau) \rightarrow \infty$ within a finite proper time.
This may or may not be accompanied by a Weyl singularity inside the
fluid; such a singularity will not occur for example if the fluid is
a Friedmann-Lema\^{\i}tre model, as the Weyl tensor is then zero
inside the fluid. \\

The Ricci tensor is zero outside the collapsing
star; it is the Weyl tensor that diverges at the spacelike
Schwarzschild singularity in the vacuum domain. Specifically, the
Kretschman scalar is
\begin{equation}  \label{eq:Kretsch}
K\ =C_{abcd}C^{abcd}=\alpha \frac{{\cal M}^{2}}{r^{6}}\;,
\end{equation}%
where $\alpha$ is a constant, so this diverges as $%
r\rightarrow 0.$\\ 

It is the spatial inhomogeneity of the matter distribution
(non-zero density inside the fluid, zero outside)\ that generates this
singularity in the conformal structure of spacetime.\\

As seen from the outside, the mass of the star never alters; it is always equal to the initial value ${\cal M}_0$:
\begin{equation}\label{eq:mass_const}
 {\cal M}_{BH} = {\cal M}_0 = \textit{const}.
\end{equation}
This will of course not be the same if matter falls into the black hole, thereby increasing it's mass; then the horizon
is a dynamic horizon \cite{AshKri02} and the laws of black hole thermodynamics \cite{Bardeen:1973gs} come into
play to characterise the resulting changes. We also note that the above picture of the black hole formation changes drastically if instead of a homogeneous dust ball we consider an inhomogeneous one with the energy density decreasing monotonically from the centre to the surface. In that case, the IMOTS at the central singularity is future outgoing and null \cite{Joshibook1, Joshibook2} and hence the central singularity is locally naked. In such cases the spacetime will cease to be future asymptotically simple \cite{HawEll73} and proofs of a number of results of black hole thermodynamics will break down. However, in our analysis here, we only concentrate on future asymptotically simple spacetimes.

\section{An Astrophysical Black Hole: The idealised picture changes}

\begin{figure}[tb!!!!]
\includegraphics[width=3.5in]{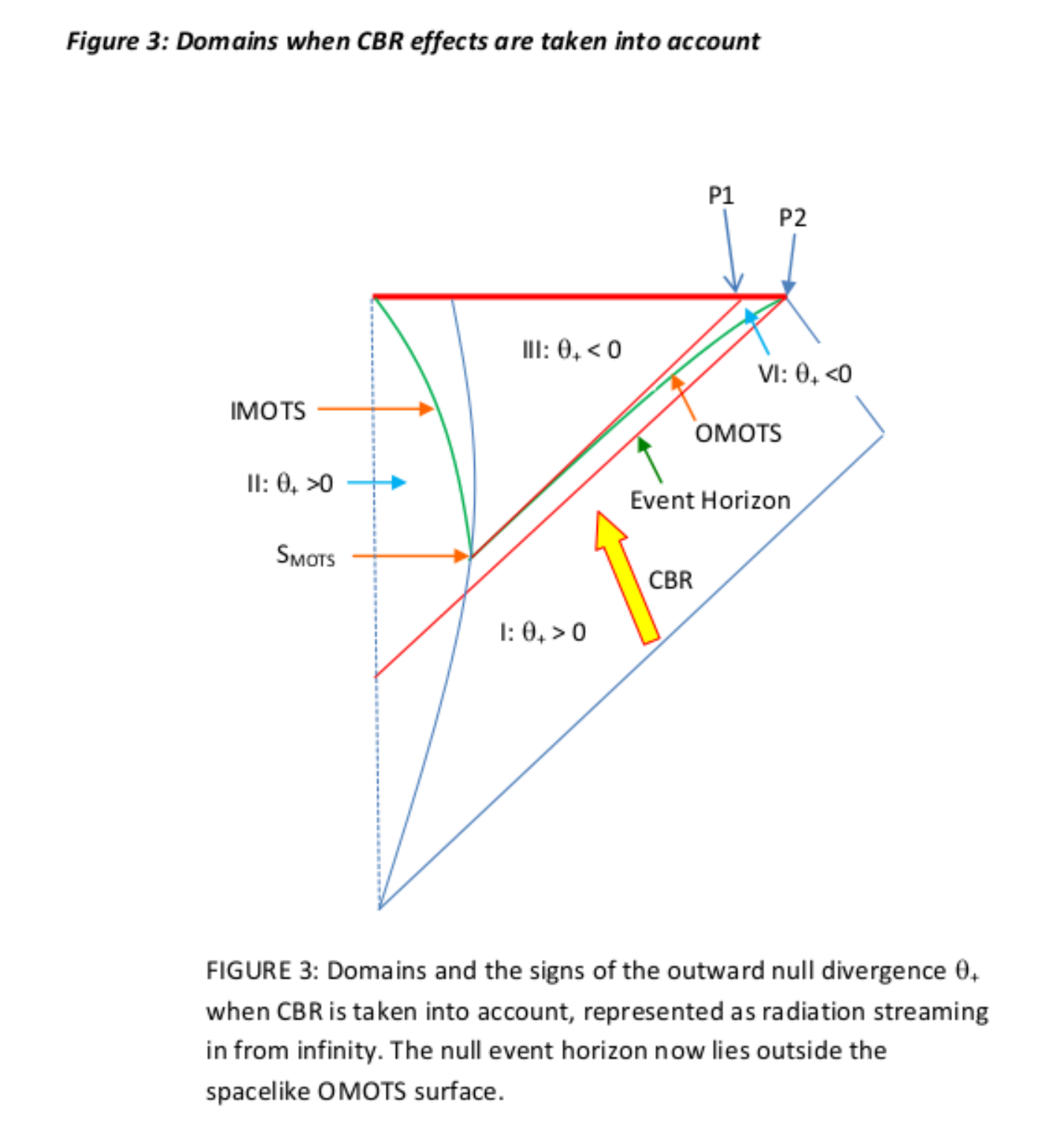}
\label{Fig3}
\end{figure}
Let us now consider a gravitational collapse scenario which is embedded in a expanding cosmology. There emerge a few crucial differences to the idealised scenario described in the previous sections. 
They are as follows:
\begin{itemize}
\item The universe is permeated by cosmic black body radiation (CBR)
radiation \ emitted from the hot big bang era in the early universe
\cite{PetUza09}, \cite{EllMaaMac12}. That blackbody background
radiation was emitted by the last scattering surface at the end of
the Hot Big Bang era in the early universe. It then propagates
through the universe with a temperature
\begin{equation}
T_{CBR}=(1+z)\,T_{CBR}|_{0}  \label{eq:CMB_Temp}\;,
\end{equation}%
where $z$ is the redshift characterising the cosmological time when
the radiation temperature is measured. It now pervades the universe
as microwave background radiation (CMB) at a temperature
\begin{equation}
T_{CBR}|_{0}=2.7K.  \label{eq:cmb_0}
\end{equation}%
In effect, this radiation reaches local systems at the present time
from an effective \textquotedblleft Finite
Infinity\textquotedblright (\cite{Ell84}, \cite{StoEll}) at about one
light year's distance that emits black body radiation at a
temperature $T_{CBR}|_{0}.$ This is the effective sky for every
local system at present. It was hotter in the past, as shown by (\ref{eq:CMB_Temp}); at very early times $T_{CBR}\simeq 4000$K and $\rho_{CBR} \simeq (1+z)^{4} \rho_{CBR}|_{0} \simeq 10^{12} \rho_{CBR}|_{0}$. The CBR radiation has the stress tensor of a perfect fluid with
positive energy density and pressure:
\begin{equation}
p_{CBR} = \rho_{CBR}/3,\,\, \rho _{CBR}=a\, T_{CBR}^{4}.
\label{eq:cmb_density}
\end{equation}
It has an entropy density given by
\begin{equation}
S_{CBR} = \frac{4}{3} a T_{CBR}^3 +b \label{eq:cmb_entropydensity}
\end{equation}
where $b$ is a constant independent of the energy $E$ and volume $V$.
\end{itemize}
The crucial outcome is that the OMOTS surface becomes spacelike \cite{Ell13} (see Figure 3). The infalling (positive density) CBR radiation  (\ref{eq:cmb_density}) increases the interior mass and makes the OMOTS surface spacelike, therefore enlarging the trapping region to include domain VI and extending the spacelike future singularity further out from P1 to P2. This moves  the globally defined event horizon outwards so that the spacelike OMOTS surface lies inside the event horizon. In fact most astrophysical black holes have an accretion disk and the rate of infalling of matter in general is much larger than the CBR. However the CBR influx rate can be considered as a lower limit to the net accretion.

\begin{figure}[htb!!!!]
\includegraphics[width=4in]{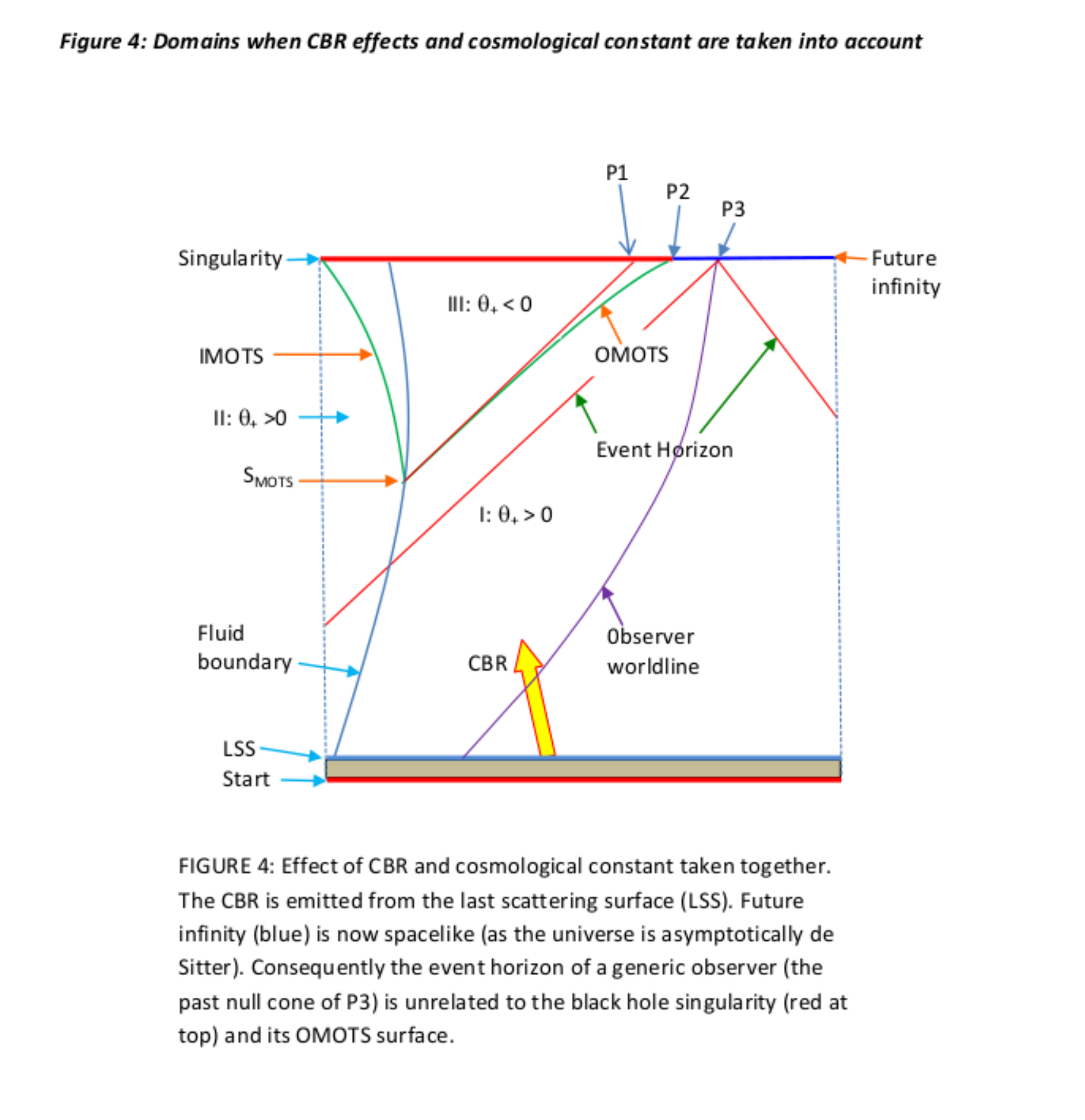}
\label{Fig4}
\end{figure}
\begin{itemize}
\item The late universe is dominated by dark energy that is consistent with existence of a small positive cosmological constant $\Lambda$. As the cosmological constant does not decay away, the late time evolution of the universe is de Sitter, and future infinity is spacelike \cite{Pen64,HawEll73} (see Figure 4). 
\end{itemize}
This means that, unlike the asymptotically flat case, there is no single event horizon for observers outside a black hole. Different cosmological event horizons (which are in the far future) occur for different observer motions that end up at different places $p_i$ on future infinity, because the event horizon for an observer ending up at $p_i$ is, by definition, the past light cone of $p_i$ \cite{Pen64,HawEll73}. The first event horizon is the past light cone of P2 (the past light cone of P1 is no longer an event horizon as it is trapped), but it is irrelevant to most observers, such as those that end up at P3. \\

Since these cosmological effects are much smaller than the thermodynamical and gravitational effects of a massive collapsing star locally, we can treat these effects as a perturbations of the idealised scenario. Let us consider that the thermodynamical variables of the CBR are $\bra{\rho_{CBR}, p_{CBR}, Q_{CBR} ,\Pi_{CBR}}$ and these are much smaller than the local gravitational tidal effects (${\cal E}$) outside a collapsing star. In this perturbed case the ratio $\sfr{\alpha}{\beta}$ for the OMOTS outside the collapsing star 
is given as 
\be\label{alph1}
\frac{\alpha}{\beta}=\frac{1+\frac23\frac{\rho_{CBR}+\Lambda}{{\cal E}}+\frac12\frac{\Pi_{CBR}}{{\cal E}}-\frac{Q_{CBR}}{{\cal E}}}{1-\frac13\frac{\rho_{CBR}+\Lambda}{{\cal E}}-\frac{p_{CBR}-\Lambda}{{\cal E}}-\frac12\frac {\Pi_{CBR}}{{\cal E}}+\frac{Q_{CBR}}{{\cal E}}}\;.
\ee
Here we note that the CBR is falling ``{\it into}" the black hole and hence we have $Q<0$. Furthermore in the background Schwarzschild spacetime ${\cal E}<0$. Using this, we find the condition for a spacelike OMOTS ($\sfr{\alpha^2}{\beta^2}<1$) is given by 
\be
\bra{\rho_{CBR}+ p_{CBR}+2|Q_{CBR}| +\Pi_{CBR}}>0\;.
\label{cond1}
\ee
The above condition is generally satisfied for CBR and hence we can easily conclude that in the presence of CBR falling in the black hole, the OMOTS is locally spacelike (see Fig. 3 for the compact description). 

Since the OMOTS is spacelike, there always exist an open set ${\cal U}$ in the vicinity of  the OMOTS from which no timelike or null trajectories can escape to infinity. To illustrate this by an example, let the instant of time for the $S_{MOTS}$ to appear in the spacetime be denoted by $\tau_0$. Let us now consider the open set ${\cal U}$ in the $[u,e]$ plane with the following closure: the curves $\Psi=0$, $t=\tau_0$ and the outgoing null curve $\nu$ from $t=\tau_0$ intersecting the curve $\Psi=0$ at some later time $t=\tau_1$ (this is possible as $\Psi=0$ is spacelike between $\tau_0$ and $\tau_1$). One can easily see that any non-spacelike trajectory from ${\cal U}$ cannot escape to infinity, the trajectories will fall back to the singularity. By construction 
this open set ${\cal U}$ 
extends to spacelike future infinity for a LCDM universe.\\

Now let us consider a realistic astrophysical Black Hole  immersed in an environment containing matter. In that case there will be
continuous in-falling  matter into the black hole and the black hole mass will vary continuously.
Now the variation of Misner-Sharp mass along $u^a$ and the preferred spatial direction $e^a$ can be calculated
using the field equations together with (\ref{mass}) as
\ba
\hat{{\cal M}}&=&\frac{1}{4K^{3/2}}\left[\phi(\rho_{CBR}+\Lambda)-\bra{\Sigma-\sfr{2}{3}\Theta}Q_{CBR}\right]\label{mhat},\\
\dot{{\cal M}}&=&\frac{-1}{4K^{3/2}}\left[\bra{\sfr{2}{3}\Theta-\Sigma}(p_{CBR}-\Lambda)+Q_{CBR}\phi\right]\label{mdot}.
\ea
Using the above equations along with the equation of horizon, we can find the evolution equation governing the mass of a black hole enclosed within the dynamical 
horizon as 
\ba
 \dot{{\cal M}}_{BH}&=&\frac{1}{4K_{BH}^{3/2}}\phi_{BH}\left[(p_{CBR}-\Lambda)+|Q_{CBR}|\right],\label{mdot1}
 \ea
 where $K_{BH}$ and $\phi_{BH}$ are evaluated on the MOTS locally.
For in-falling CBR  we have $p_{CBR}>0$. Also as already stated earlier, in the interior and exterior spacetime we have $\phi\ge0$ . Hence 
we see as the CBR falls in the black hole, its mass ${\cal M}_{BH}$ increases monotonically. {As the CBR becomes weaker and weaker in far future the thermodynamic terms will be less than the cosmological constant $\Lambda$. When $\Lambda$ dominates, the solution tends to a Schwarzschild de Sitter solution where we have  $\sfr{\alpha}{\beta}=1$ and the OMOTS continuously tends to a null surface from a spacelike one, with $\phi_{BH}\rightarrow0$ till the future spacelike infinity. Hence during this phase near the future spacelike infinity $\dot{{\cal M}}_{BH}\rightarrow 0$ and the black hole mass will asymptotically reach a constant value.} 
\section{Discussions: Consequences for Hawking Radiation}

The results obtained in the previous section for location of horizons of realistic astrophysical black holes immersed in a LCDM cosmology with CBR, may have important consequences on the Hawking radiations from these black holes. We return now to the key questions raised in the introductory section.

\subsubsection{Local or Global:}
As Fig 4 depicts, we would like to emphasise that there are no global event horizons in an asymptotically de-Sitter Universe. The event horizon for each particle is different, being the past null cone of it's world line at future spacelike infinity.Even the particle P1, whose worldline ends
at a singularity, has every right to refer to it's past lightcone also as a future event horizon. In the context of string theory these null boundaries are sometimes referred as {\it observer horizons}. Hence there exist an infinite number of world lines that will never see Hawking radiation from a black hole, as this radiation (if any) will lie outside their cosmological event horizon. Therefore it makes no sense to talk about Hawking radiation from a global event horizon in this context. Although there is a notion of this happening in the Gibbons-Hawking vacuum, each observer sees radiation coming from their own personal event horizon. But the relation of such radiation to a black hole horizon (which may lie entirely outside the observer's event horizon) is yet to be deciphered properly.  

We can argue that there is an inmost event horizon associated with the black hole, so one might propose radiation is associated with this non-locally determined surface. We would like to clarify here that this inmost horizon is the part of the event horizon that forms before the incoming CBR shell crosses the horizon. However from Fig. 1, it can be seen that this event horizon forms even before parts of the spacetime become trapped. 
It seems unlikely that the event horizon in such a regular and untrapped epoch will start radiating as there is no local occurrence there that could lead to this happening. Hence it is probable the radiation must come from the vicinity of the boundary of the trapped region, that is, from near an apparent horizons (a MOTS).  

\subsubsection{Timelike or spacelike:}
If the radiation is locally emitted \cite{Vis01}, it will be associated with either the IMOTS surface or the OMOTs surface.
As was suggested in the introduction,  if the calculation for Hawking radiation is based on the idea of tunnelling \cite{ParWil00}, then it will only be applicable if the MOTS surface is timelike or null. If it is spacelike, then a tunnelling viewpoint is simply not applicable. Hence in the scenario discussed in this paper only the timelike IMOTS could radiate; and any such radiation emitted near the IMOTS will necessarily fall into the singularity and so be shielded from outside observers \cite{Ell13}. 

However that conclusion depends on the particle picture, which can be called into question. Consider then possible emission near the OMOTS surface. 
As we proved in the previous sections, the CBR makes the OMOTS spacelike, which ensures the existence of an open set in the vicinity of the OMOTS from which no timelike or null trajectories can escape to infinity. Hence if any Hawking radiation is generated from this open set it is bound to fall back into the black hole rather than escaping to infinity. This suggests that there may be a remnant mass of any black hole at future infinity, and the black hole will not evaporate entirely away.  Quantum Field Theory calculations suggest radiation is emitted from the vicinity of the OMOTS surface, rather than just from the surface itself, so on this picture a small amount of radiation that is emitted from outside the open set described above can in principle escape to infinity. However existence of such open set would necessarily imply a part of the radiation would fall back to the black hole and therefore the entire black hole will not totally evaporate away.

This conclusion would however be wrong if the backreaction from ingoing negative density Hawking radiation overcame the CBR effect and made the OMOTS surface timelike. Can this occur?  If we consider the usual particle picture of Hawking radiation from an open set ${\cal U}$ described above between two closely spaced interval $\tau_1$ and $\tau_2$, we can easily see that both the positive energy particle and the negative energy particle will enter subset of the trapped region between $\tau_1$ and $\tau_2$. Hence there will be no net change in the average energy (mass) of the black hole due to Hawking radiation from this open set. This can be better understood in the context of local black hole mass. As we have seen, in-falling CMBR increases the black hole mass monotonically, and that implies the temperature of the horizon decreasing monotonically. We know at the present epoch the temperature of CBR is much greater than the Hawking temperature of a solar mass black hole. {From equation (\ref{mdot1}) we can easily see that the mass of the black hole will go on increasing (and hence the Hawking temperature will go on decreasing) till $p_{CMB}$ and $|Q_{CMB}|$ becomes smaller than the cosmological constant. From this point in far future to the future spacelike infinity the black hole mass asymptotically tend to a constant value with the OMOTS asymptotically becoming null from a spacelike surface but never actually being null: it is always spacelike. Hence much of the emitted Hawking radiation will be trapped, as the OMOTS surface will remain inside the black hole event horizon.  In the thermodynamic context, black holes not radiating to a hotter background makes perfect sense.}

This argument is suggestive but not conclusive: to obtain a definitive result, we must calculate the expectation value of the stress tensor in the present context (work under way). However insofar as the particle picture is valid, the above conclusion seems inevitable.   

We would like to clarify here that our conclusion is different from that in Bardeen's paper \cite{Bar14} \textit{inter alia} because his   
picture of the horizon geometry is different from ours. In that paper he used as a classical background a somewhat artificial construction of a singularity free black hole \cite{Hay06,Fro14} leading to the existence of a particle creation surface ``outside'' the black hole with zero energy density but non-zero surface stress. The existence of such a surface is unphysical, and we argue that it cannot develop for a realistic astrophysical black hole. It is the use of this unphysical equation of state that leads to an understanding of the inner horizon (see Fig 2 in \cite{Bar14}) different than that presented here. 
 
\subsubsection{Vacuum or fluid:} 
 It is quite interesting that most of the existing literature on Hawking radiation from a black hole talks about vacuum excitations near the event horizon. Very few works discuss the effect of a time varying collapsing matter field on the particle creation; however there may well be particle emission there too \cite{Par69,Haw73,BirDav84}.  
 
 In \cite{BirDav84} it is argued that if particles created by vacuum excitation in the exterior of an collapsing star travels through the star, then the difference in blueshift of a particle falling in from the redshift of the same particle coming out 
 becomes important in the vicinity of the horizon. This difference in particle energy manifests in a diminishing energy of the collapsing star, which might stop the formation of singularities and trapped surface \cite{Mer14}. In that case all the above argument is called into question.
 
 Whether such effects can still occur in the cosmological context discussed here, where a spacelike OMOTS surface occurs inside the event horizon, is an open question, and is worthy of future investigation.

 \begin{acknowledgments}
 GFRE thanks Paul Davies, Don Page, Tim Clifton, and particularly Malcolm Perry for helpful interactions that have shaped much of this work, and Hadi and Mahdi Godazgar  and Laura Mersini Houghton for useful discussions on the topic at Great Brampton House in March this year. We thank Charles Hellaby, Alton Kesselly and Javad Firouzjaee for helpful discussions in Cape Town, and Tim Clifton, Malcolm Perry, and Naresh Dadich for comments on an earlier draft that have helped improve this paper.
 
  AH and RG are supported by National Research Foundation (NRF), South Africa, and GFRE receives support from the NRF and from the UCT research committee. SDM acknowledges that this work is based on research supported by the South African Research Chair Initiative of the Department of
Science and Technology and the National Research Foundation.

We thank the anonymous referee for his valued comments on this paper.
 \end{acknowledgments}

\end{document}